\documentclass[showpacs,10pt,twocolumn,prb]{revtex4-1}

\usepackage{amsmath}
\usepackage{amssymb}
\usepackage{graphics}
\usepackage{epsfig}
\usepackage{CJK}
\usepackage{color}

\begin{document}

\begin{CJK*}{GBK}{Song}

\title{Thermoelectric studies of Ir$_{1-x}$Rh$_x$Te$_2$ (0 $\leqslant x \leqslant $ 0.3)}
\author{Yu Liu,$^{1}$ Hechang Lei,$^{1,\ast}$ Kefeng Wang,$^{1,\dag}$ Milinda Abeykoon,$^{2}$ J. B. Warren,$^{3}$ Emil Bozin,$^{1}$ and C. Petrovic$^{1}$}
\affiliation{$^{1}$Condensed Matter Physics and Materials Science Department, Brookhaven National Laboratory, Upton, New York 11973, USA\\
$^{2}$Photon Science Division, National Synchrotron Light Source II, Brookhaven National Laboratory, Upton, New York 11973, USA\\
$^{3}$Instrumentation Division, Brookhaven National Laboratory, Upton, New York 11973, USA}
\date{\today}

\begin{abstract}
We report thermoelectric properties of Ir$_{1-x}$Rh$_x$Te$_2$ ($0 \leqslant x \leqslant 0.3$) alloy series where superconductivity at low temperatures emerges as the high-temperature structural transition ($T_s$) is suppressed. The isovalent ionic substitution of Rh into Ir has different effects on physical properties when compared to the anionic substitution of Se into Te, in which the structural transition is more stable with Se substitution. Rh substitution results in a slight reduction of lattice parameters and in an increase of number of carriers per unit cell. Weak-coupled BCS superconductivity in Ir$_{0.8}$Rh$_{0.2}$Te$_2$ that emerges at low temperature ($T_c^{zero}$ = 2.45 K) is most likely driven by electron-phonon coupling rather than dimer fluctuations mediated pairing.
\end{abstract}

\maketitle
\end{CJK*}

\section{Introduction}
The complex and unusual structural and physical properties of layered transition metal dichalcogenides (TMDCs) have been investigated for several decades, in part due to the competing orders such as charge density wave (CDW) order and superconductivity (SC).\cite{Wilson1,Wilson2,Castro,Valla,LiLJ} The interplay of competing orders is the fundamental question in these systems. The CDW state can be tuned into SC via intercalation, substitution, pressure or electric-field effect.\cite{LiLJ,Liu,Sipos} Usually, there is a dome-like phase diagram, i.e., the CDW transition temperature $T_{CDW}$ decreases when the superconducting critical temperature $T_c$ increases. This indicates that CDW and SC compete, however details of this competition are material-dependent and not well understood.\cite{LiLJ,Liu,Sipos,Gabovich,Liu1,Liu2}

The discovery of superconductivity in Pt, Pd, and/or Cu substituted/intercalated CdI$_2$-type IrTe$_{2}$ with its $T_c$ up to $\sim$ 3 K has triggered a resurgence of interest in this field.\cite{Pyon,Yang JJ,Kamitani} IrTe$_2$ is a TMDC that undergoes a structural transition at $\sim$ 270 K from trigonal P$\bar3$m1 symmetry to triclinic P$\bar1$.\cite{Matsumoto,Pascut,Toriyama} The transition is accompanied
by partial Ir-Ir dimerization associated with substantial structural distortions.\cite{Pascut,Toriyama} Transmission electron microscope, photoemission, electron diffraction and tight-binding electronic structure calculations all revealed a superstructure with $\textbf{q}$ = (1/5, 0, -1/5) modulation vector associated with the low temperature phase and initially ascribed to an orbitally driven Peierls instability.\cite{Yang JJ,Ootsuki2} However, results from nuclear magnetic resonance (NMR), angle-resolved photoemission spectroscopy (ARPES), optical conductivity and scanning tunneling microscopy/spectroscopy measurements brought to doubt the conventional CDW instability scenario, and instead suggested that the transition is due to the reduction of the kinetic energy of Te $p$ bands.\cite{Fang AF,Mizuno,Ootsuki,LiQing} In addition, theoretical calculations suggest that the structural transition is mainly caused by the evolution of Te $p$ bands rather than the instability of Ir $d$ bands which results in a reduction of the kinetic energy of the electronic system.\cite{Fang AF, Kamitani} On the other hand, with Pt, Pd, Cu substitution/intercalation, the structural transition is quickly suppressed and superconductivity appears at low temperature, indicating the competition between the two order parameters.

Isovalent substitution is an effective way to further inform the discussion about the origin of transition and induced superconductivity. It is similar to pressure because it changes the structural parameters, ionic size and electronegativity of atoms in the unit cell without introducing extra carriers, thus affecting electronic structure and vibrational properties. Previous studies have shown that the structural transition at high temperature is enhanced while the superconducting transition is suppressed by either hydrostatic or chemical pressure.\cite{Kiswandhi,Oh,Ivashko} The latter results are associated with stabilization of polymeric Te-Te bonds by replacing Te with the more electronegative Se. This is different from other TMDCs where pressure suppresses the CDW state and enhances the SC.\cite{Liu, Sipos} Isovalent Rh substitution in IrTe$_2$ also induces SC,\cite{Kudo} however little is known about the nature of superconducting state.

In this work, we report the thermoelectric properties of Ir$_{1-x}$Rh$_x$Te$_2$ ($0\leqslant x\leqslant 0.3$). Our results give a slight reduction of lattice parameters and an increase of number of carriers per unit cell. With suppression of the high-temperature structural transition, superconducting state that emerges at low temperature in Ir$_{0.8}$Rh$_{0.2}$Te$_2$ is weak-coupled BCS, suggesting conventional electron-phonon mechanism. This is consistent with recently reported absence of nanoscale dimer fluctuations in Rh,Pt-substituted IrTe$_2$ and argues against the dimer fluctuations mediated exotic superconductivity with singlet-triplet pairing.\cite{YuRunze,Ootsuki3}

\section{Experiment}

Polycrystalline samples of Ir$_{1-x}$Rh$_x$Te$_2$ ($0\leqslant x\leqslant 0.3$) were synthesized using solid-state reaction method as described previously.\cite{Yang JJ} The structure was characterized by powder x-ray diffraction (XRD) in transmission mode at the X7B beamline of the National Synchrotron Light Source (NSLS) at Brookhaven National Laboratory. Data were collected using 0.5 mm$^2$ monochromatic beam of $\sim$ 38 keV ($\lambda \sim$ 0.3916 ${\AA}$) at 300 K. A Perkin Elmer 2D detector was placed orthogonal to the beam path 376.4 mm away from the sample. Sample was loaded in a polyamide capillary 1 mm in diameter and mounted on a goniometer head. The data were collected up to Q = 4$\pi$sin$\theta$/$\lambda$ = 12 ${\AA}$$^{-1}$. The average stoichiometry was determined by energy-dispersive x-ray spectroscopy (EDX) in a JEOL JSM-6500 scanning electron microscope (SEM). The specific heat was measured on warming procedure between 1.95 and 300 K by the heat pulse relaxation method using a QD physical property measurement system (PPMS) with sample mass of 20 $\sim$ 30 mg. The electrical resistivity ($\rho$), Seebeck coefficient ($S$), and thermal conductivity ($\kappa$) were measured on a QD PPMS using the thermal transport option (TTO) with standard four-probe technique. Continuous measuring mode was used. The maximum heat powder and period were set as 50 mW and 1430 s along with the maximum temperature rise of 3$\%$. The drive current and frequency for resistivity is 1 mA and 17 Hz. The sample dimensions were measured by an optical microscope Nikon SMZ-800 with 10 $\mu$m resolution.

\section{Results and Discussions}

\begin{figure}
\centerline{\includegraphics[scale=1]{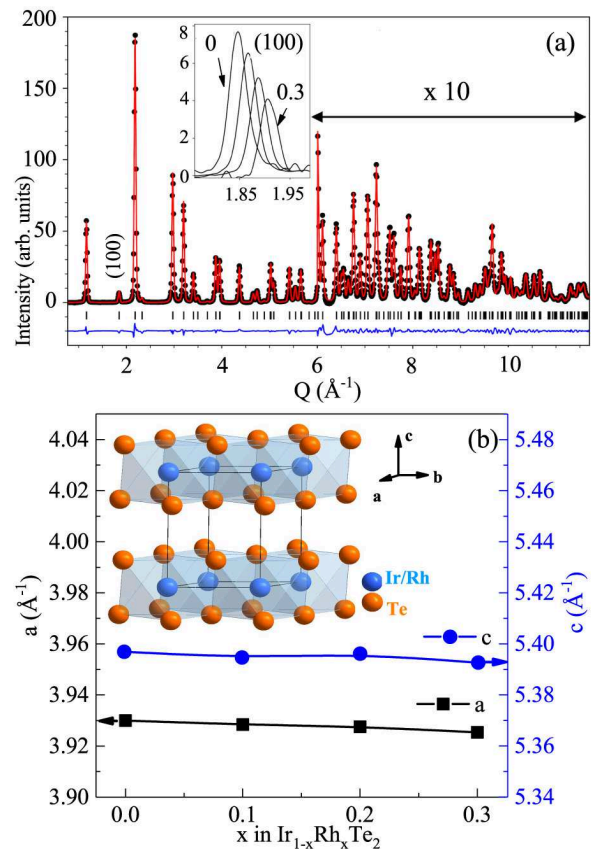}}
\caption{(a) The Rietveld refinement of the background subtracted IrTe$_2$ synchrotron powder x-ray diffraction up to Q $\sim$ 12 {\AA}$^{-1}$. Plots show the observed (dots) and calculated (red solid line) powder patterns with a difference curve. The vertical tick marks represent Bragg reflections in the P$\bar3$m1 space group. Inset shows the evolution of the normalized intensity of (100) Bragg reflection with increasing $x$ in Ir$_{1-x}$Rh$_x$Te$_2$. (b) Unit cell parameters as a function of Rh substitution up to $x$ = 0.3. Inset shows crystal structure of Ir$_{1-x}$Rh$_x$Te$_2$ with Ir/Rh sites marked in blue and Te sites marked in orange.
}
\end{figure}

The EDX measurement confirmed Ir$_{1-x}$Rh$_x$Te$_2$ stoichiometry and nominal Ir/Rh ratio within up to 10$\%$ experimental error. Rietveld powder diffraction analysis was carried out on data sets obtained from the raw 2D diffraction data integrated and converted to intensity versus $Q$ using the software Fit2d where $Q$ is the magnitude of the scattering vector.\cite{Hammersley} The refinement was performed using GSAS/EXPGUI modeling suite.\cite{Larson,Toby} We used the room temperature CdI$_2$ prototype structure and trigonal symmetry (P$\bar3$m1, 1T phase).\cite{Young,Hockings} Figure 1(a) shows fits to the data with no impurity peaks present. Rietveld analysis produced excellent fits to the data up to a high Q, suggesting high purity of samples and high quality of the XRD data.

Ir$_{1-x}$Rh$_x$Te$_2$ crystalizes in a layered structure [inset in Fig. 1(b)]. There is a large number of compounds crystalizing in this structure, especially TMDCs such as MX$_2$ (M = Ti, Ta, or Nb, X = S, Se, or Te). In this structure, the edge-sharing Ir/Rh-Te octahedra form Ir/Rh-Te layers in the $ab$ plane, resulting in the network of equilateral triangles populated by Ir ions. The sandwich-like Te-Ir/Rh-Te layers stack along the $c$ axis with Te-Te bonds instead of weak van der Waals gap which has been often observed in TMDCs.\cite{Yang JJ, Lee1, Pettenkofer} Although the interlayer interaction might be stronger than in typical TMDCs, there are still some ions that can be intercalated between Te-Ir/Rh-Te layers, such as Pd and Cu.\cite{Yang JJ, Kamitani} The intercalation usually results in the increase of the $c$ axial lattice parameter.\cite{Yang JJ, Pyon} On the other hand, for the Pt/Pd substitution, the $c$ axial lattice parameter decreases with doping. The $a$ and $c$ axial lattice parameters of Ir$_{1-x}$Rh$_x$Te$_2$ ($0\leqslant x\leqslant 0.3$) decrease only weakly with Rh substitution [Fig. 1(b)], which might be partially due to the similar ionic radii of Ir and Rh.

\begin{figure}
\centerline{\includegraphics[scale=1]{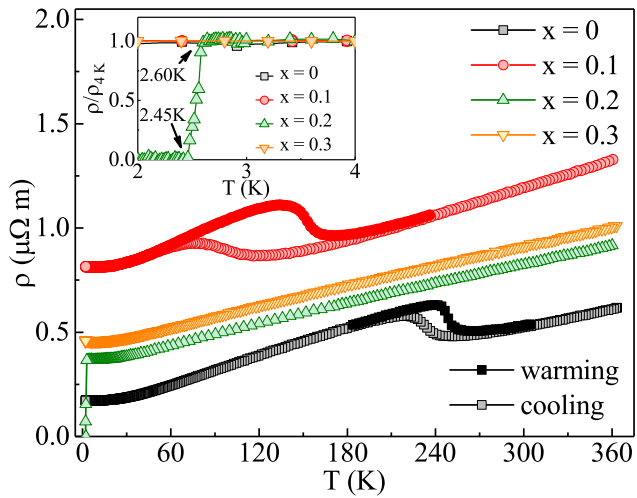}}
\caption{Temperature dependence of electrical resistivity for Ir$_{1-x}$Rh$_{x}$Te$_{2}$ (0 $\leqslant x \leqslant $ 0.3). Inset shows the enlarged low temperature part of 2 $\sim$ 4 K (normalized at 4 K).}
\end{figure}

Temperature-dependent resistivity of pure IrTe$_{2}$ ($x = 0$) shows metallic behavior with a significant thermal hysteresis around $\sim$ 250 K, as depicted in Fig. 2, which has been ascribed to a structural transition from the trigonal P$\bar{3}$m1 space group to triclinic symmetry P$\bar{1}$.\cite{Matsumoto,Pascut,Toriyama} However, the origin of this structural transition is still disputed. Electron diffraction suggests that the structural transition is driven by charge-orbital density wave with a wave vector of $q = \{1/5, 0, -1/5\}$.\cite{Yang JJ} On the other hand, the NMR experiment does not provide the evidence for CDW and the optical spectroscopic as well as ARPES measurements do not observe the gap that would correspond to the CDW state near the Fermi level.\cite{Mizuno, Fang AF, Ootsuki} Additionally, the theoretical calculation suggests that the structural transition is mainly caused by the evolution of Te $p$ bands rather than the instability of Ir $d$ bands, which results in a reduction of the kinetic energy of the electronic system.\cite{Fang AF, Kamitani} With Rh substitution, the hysteresis becomes broad and shifts to lower temperature. This anomaly disappears and the superconductivity (above 2 K) emerges in Ir$_{0.8}$Rh$_{0.2}$Te$_2$ with the transition temperatures $T_c^{onset}$ = 2.60 K and $T_c^{zero}$ = 2.45 K [inset in Fig. 2], comparable to those with other dopants or intercalating agents.\cite{Yang JJ, Pyon, Kamitani}

Figure 3 shows the temperature dependence of Seebeck coefficient $S(T)$ for Ir$_{1-x}$Rh$_x$Te$_2$ ($0\leqslant x\leqslant 0.3$). The sign of $S(T)$ for all samples in the whole temperature range is positive, indicating the hole-type carriers. The $S(T)$ of pure IrTe$_2$ ($x = 0$) shows a reduction at (220-260) K, which reflects the reconstruction of the Fermi surface across the structural transition $T_s$; i.e., the dimer formation below the $T_s$ is likely related to partial localization of some hole-type carriers at $T_s$.\cite{Pascut,YuRunze} With Rh substitution, the hysteresis becomes broad and shifts to lower temperature for $x$ = 0.1, and disappears for $x$ = 0.2 and 0.3. Rh doping suppresses not only structure transition but also the corresponding Fermi surface reconstruction, offering evidence that the structural and electronic transitions are closely coupled. The high-temperature transition is not present for $x$ = 0.2 sample, which is a much larger substitution content when compared to Pt, Pd, or Cu substitution/intercalation where only several percent ($<$ 5\%) will suppress the $T_s$ completely.\cite{Yang JJ, Pyon, Kamitani} This possibly reflects smaller atomic radii difference and isovalent character of substitution of Rh for Ir as compared to Pt substitution.

\begin{figure}
\centerline{\includegraphics[scale=1]{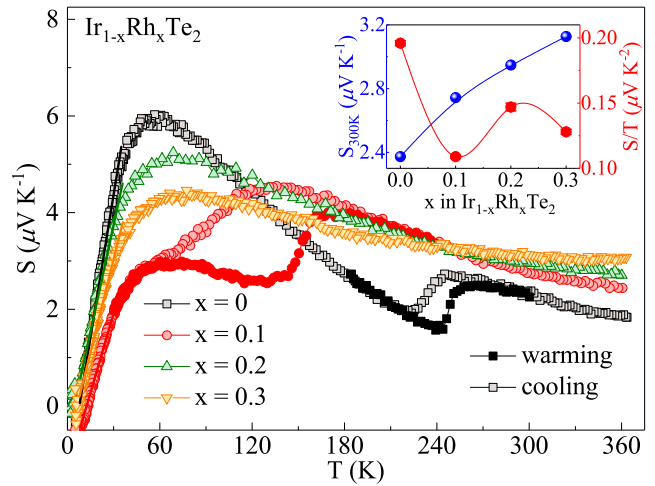}}
\caption{Temperature dependence of Seebeck coefficient for Ir$_{1-x}$Rh$_x$Te$_2$ (0 $\leqslant x \leqslant $ 0.3). Inset shows the evolution of $S_{300K}$ (balls) and $S/T$ (circles) as a function of Rh content $x$.}
\end{figure}

Thermopower gradually increases with decreasing temperature in all investigated samples. Below about (50-60) K the diffusive Seebeck response dominates. As we know, in a metal with dominant single-band transport, the Seebeck coefficient could be described by the Mott relationship, $S = \frac{\pi^2k_B^2T}{3e}\frac{N(\varepsilon_F)}{n}$, where $N(\varepsilon_F)$ is the density of states (DOS), $\varepsilon_F$ is the Fermi energy, $n$ is the carrier density, $k_B$ is the Boltzman constant and $e$ is the absolute value of electronic charge.\cite{TE} The room temperature value of $S_{300K}$ increases with increasing $x$ (inset in Fig. 3), implying the decrease in hole-type carriers with Rh substitution in Ir$_{1-x}$Rh$_x$Te$_2$ and the absence of their reduction due to transition in $x$ = 0.2 and 0.3 samples. Interestingly, the $S/T$ value determined by linear fitting below 35 K exhibit a nonmonotonic trend with $x$, with $x$ = 0.1 composition close to dimer/SC boundary being affected by the suppressed structural/electronic transition close to the observed Seebeck peak. The $S/T$ value is associated with the strength of electron correlations, calling for more theoretical study in-depth along with ARPES measurement. For the broad $S(T)$ peak around (50-60) K, generally, the phonon drag contribution to $S(T)$ gives $\sim$ $T^3$ dependence for $T \ll \Theta_D$, $\sim 1/T$ for $T \gg \Theta_D$, and a peak feature for $T \sim$ $\Theta_D/5$. The Debye temperature $\Theta_D$ of pure IrTe$_2$ ($x$ = 0) is $\sim$ 215(1) K, decreasing to $\sim$ 198(1) K for $x$ = 0.3 (as discussed below). Therefore, the peak position of $S(T)$ from phonon drag should be at $\sim$ 43.0(2) K for $x$ = 0, decreasing to 38.8(2) for $x$ = 0.3, which is lower than the experimentally observed values. This further suggests the dominance of electronic diffusion over the phonon drag effect in thermoelectricity of Ir$_{1-x}$Rh$_x$Te$_2$ ($0\leqslant x\leqslant 0.3$).

\begin{figure}
\centerline{\includegraphics[scale=1]{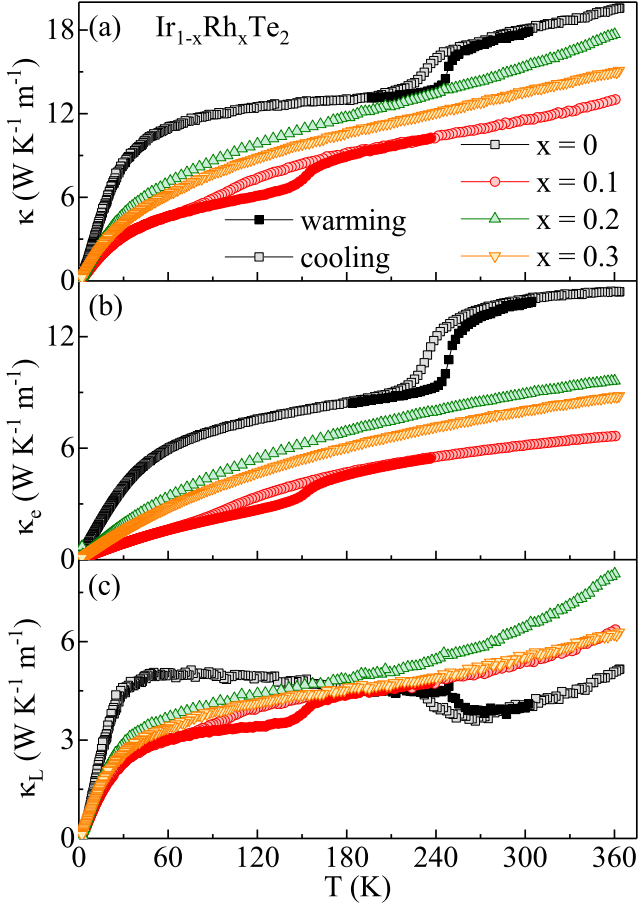}}
\caption{Temperature dependence of (a) total thermal conductivity $\kappa$, (b) electronic part $\kappa_e$, and (c) phonon term $\kappa_L$ for Ir$_{1-x}$Rh$_x$Te$_2$ (0 $\leqslant x \leqslant $ 0.3).}
\end{figure}

Temperature-dependent total thermal conductivity $\kappa$ for Ir$_{1-x}$Rh$_x$Te$_2$ (0 $\leqslant x \leqslant $ 0.3) [Fig. 4(a)] shows that the $\kappa(T)$ drop that corresponds to the $T_s$ is also clearly observed. Generally, $\kappa = \kappa_e + \kappa_L$, consists of the electronic part $\kappa_e$ and the phonon term $\kappa_L$. The $\kappa_L$ can be obtained by subtracting the $\kappa_e$ part calculated from the Wiedemann-Franz law $\kappa_e/T = L_0/\rho$, where $L_0$ = 2.45 $\times$ 10$^{-8}$ W $\Omega$ K$^{-2}$ and $\rho$ is the measured resistivity. The estimated $\kappa_e$ and $\kappa_L$ are depicted in Figs. 4(b) and 4(c). It shows that $\kappa_e$ is larger and comparable to $\kappa_L$, and confirms the substantial effect of crystal structure changes on heat-carrying phonons. Phonon-related thermal conductivity in pure IrTe$_2$ ($x$ = 0) is slightly increased below the $T_s$ whereas Rh substitution brings an increase in $\kappa_L$ in the temperature range above the $T_s$. The $\kappa_L$ of pure IrTe$_{2}$ ($x$ = 0) features a broad maximum around (50-60) K, which is significantly suppressed with Rh doping at Ir sites. This is mostly contributed by the Rh/Ir doping disorder enhanced point defects scattering. It should be noted that the $x$ = 0.1 sample shows the smallest value of $\kappa(T)$ as well as $S(T)$ at low temperature. The $\kappa(T)$ and $S(T)$ share the same tendency, suggesting increased possibility for phonon drag with Rh substitution $x$.

\begin{figure}
\centerline{\includegraphics[scale=1]{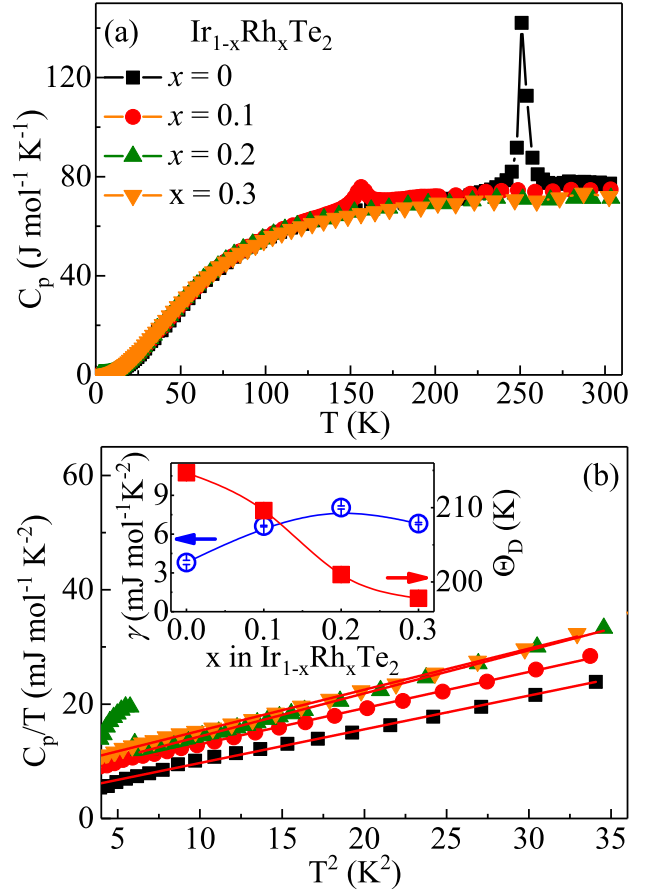}}
\caption{(a) Temperature dependence of specific heat $C_p$ for Ir$_{1-x}$Rh$_x$Te$_2$ (0 $\leqslant x \leqslant $ 0.3). (b) The low temperature specific heat divided by temperature $C_p/T$ as a function of $T^2$ in zero field. The solid curve represents the fittings using $C_p/T=\gamma+\beta T^2$. Inset shows the evolution of Debye temperature $\Theta_D$ and electronic specific heat coefficient $\gamma$.}
\end{figure}

Figure 5(a) shows the specific heat of Ir$_{1-x}$Rh$_x$Te$_2$ (0 $\leqslant x \leqslant $ 0.3) between 1.95 and 300 K. For pure IrTe$_2$ ($x$ = 0), there is a peak at 251 K, corresponding to the anomaly in thermopower and thermal conductivity. The peak shape indicates that it is a first-order transition, consistent with the reported value in the literature.\cite{Fang AF} With Rh doping, the peak shifts to 156 K for $x$ = 0.1 and the intensity of peak also becomes weaker and less sharp than that in pure one, indicating a possible change from the first- to the second-order transition near the $T_s$.\cite{Kudo} The specific heat $C_p(T)$ at low temperature above $T_c$ can be well fitted by using $C_p/T=\gamma+\beta T^2$ [Fig. 5(b)]. The evolution of the obtained $\gamma$ and derived Debye temperature $\Theta_D$ from $\beta$ using the relation $\Theta_D=(12\pi^4NR/5\beta)^{1/3}$, where $N$ = 3 is the number of atoms per formula unit and $R$ is the gas constant, are plotted in inset of Fig. 5(b). The electronic specific heat of IrTe$_2$ is $\gamma$ $\sim$ 3.81 mJ/mol-K$^2$, close to previously reported values.\cite{Pyon, Fang AF} With the increase of the Rh content, the value of $\gamma$ reaches a maximum of 8.06 mJ/mol-K$^2$ for $x$ = 0.2 and decreases to 6.82 mJ/mol-K$^2$ for $x$ = 0.3. This is similar to the Pt doping where $\gamma$ starts to decrease when $x\geq$ 0.04, ascribed to the decrease of DOS of IrTe$_2$ above the Fermi level and the upward Fermi level shift due to the partial substitution of Pt for Ir.\cite{Pyon}

The electronic specific heat $C_e=\frac{\pi^2}{3}k_B^2TN(\varepsilon_F)$, where $N(\varepsilon_F)$ is the DOS, $\varepsilon_F$ is the Fermi energy, and $k_B$ is the Boltzman constant. Taken into consideration the Mott relationship $S = \frac{\pi^2k_B^2T}{3e}\frac{N(\varepsilon_F)}{n}$, thermopower probes the specific heat per electron: $S = C_e/ne$, where the units are V K$^{-1}$ for $S$, J K$^{-1}$ m$^{-3}$ for $C_e$, and m$^{-3}$ for $n$, respectively. However, it is common to express $\gamma = C_e/T$ in J K$^{-2}$ mol$^{-1}$ units. In order to focus on the $S/C_e$ ratio, let us define the dimensionless quantity, $q=\frac{S}{T}\frac{N_Ae}{\gamma}$ where $N_A$ the Avogadro number, gives the number of carriers in the unit cell (proportional to $1/n$).\cite{Behnia} The constant $N_Ae$ = 9.6 $\times$ 10$^5$ C mol$^{-1}$ is also called the Faraday number. We observe reduction of $q$ from 0.49(1) for pure IrTe$_2$, to 0.16(1) for $x$ = 0.1, and 0.18(1) for $x$ = 0.2 and 0.3. This implies an increase of carrier concentration as Rh enters the lattice.

Then we discuss the superconducting state observed in Ir$_{1-x}$Rh$_{x}$Te$_2$ ($x$ = 0.2). According to the McMillan formula for electron-phonon mediated superconductivity,\cite{McMillan} the electron-phonon coupling constant $\lambda$ can be deduced by
\begin{align*}
T_c=\frac{\Theta_D}{1.45}\exp[-\frac{1.04(1+\lambda)}{\lambda-\mu^{\ast}(1+0.62\lambda)}],
\end{align*}
where $\mu^{\ast}\approx$ 0.13 is the common value for Coulomb pseudopotential. By using $T_c$ = 2.45(1) K and $\Theta _D$ = 201(1) K in Ir$_{0.8}$Rh$_{0.2}$Te$_2$, we obtain $\lambda \approx$ 0.59(2), a typical value of weak-coupled BCS superconductor. The specific heat jump at $T_c$, $\Delta$C$_{es}$/$\gamma T_c\approx$ 1.18, is somewhat smaller than the weak coupling value 1.43.\cite{McMillan} All these results indicate that Ir$_{0.8}$Rh$_{0.2}$Te$_2$ is a weak-coupled BCS superconductor, similar to Ir$_{0.95}$Pd$_{0.05}$Te$_2$.\cite{YuDJ} Recent x-ray pair distribution function approach reveals that the local structure of Ir$_{0.8}$Rh$_{0.2}$Te$_2$ could be well explained by a dimer-free model, as well as in Ir$_{0.95}$Rh$_{0.05}$Te$_2$, ruling out the possibility of there being nanoscale dimer fluctuations mediated superconducting pairing.\cite{YuRunze}

\section{Conclusion}

In summary, we have investigated thermal transport properties of Ir$_{1-x}$Rh$_x$Te$_2$ (0 $\leqslant x \leqslant $ 0.3). The decrease of thermpower in pure IrTe$_2$ at $T_s$ implies that dimer formation is related to partial localization of hole-type carriers below the structural transition. The isovalent Rh substitution in Ir$_{1-x}$Rh$_x$Te$_2$ results in an increase of number of carriers per unit cell as Rh enters the lattice and weak-coupling BCS superconductivity. Further investigation of stripe phases existence in Ir$_{1-x}$Rh$_x$Te$_2$ using real-space resolving methods would be useful for addressing the microscopic connection of phonon-related Cooper pairing on the border of exotic structural transitions.\cite{DaiJixia, MaurererT}

\section{Acknowledgements}

We thank the X7B at the NSLS Brookhaven National Laboratory for the use of their equipment. Work at Brookhaven is supported by the U.S. DOE under Contract No. DE-SC0012704.

$^{\ast}$ Present address: Department of Physics and Beijing Key Laboratory of Opto-electronic Functional Materials and Micro-nano Devices, Renmin University of China, Beijing 100872, People's Republic of China.

$^{\dag}$ Present address: Center for Nanophysics and Advanced Materials, Department of Physics, University of Maryland, College Park, MD 20742, USA.

\end{document}